\documentclass{ws-procs9x6-cpt16}
\begin{document}

\newcommand{\refeq}[1]{(\ref{#1})}
\def\etal {{\it et al.}}

\title{Testing Lorentz and CPT Symmetries in Penning Traps}

\author{Yunhua Ding}

\address{Physics Department, Indiana University\\
Bloomington, IN 47405, USA}

\begin{abstract}
A modified Dirac equation with general Lorentz- and CPT-violating operators
in the electromagnetic field is studied. 
Constraints on and possible sensitivities to Lorentz-violating coefficients 
in the nonminimal sector up to mass-dimension six can be obtained 
by analyzing Penning-trap results involving anomaly frequencies.
\end{abstract}

\bodymatter

\section{Introduction}
Lorentz and CPT symmetries are fundamental in the Standard Model,
which is tremendously successful in describing nature 
in both theoretical and experimental aspects. 
However,
these symmetries could be violated from the process of spontaneous breaking
in the underlying theory including quantum gravity,
such as strings.\cite{string}
The general framework characterizing such violations is the Standard-Model Extension (SME), 
which incorporates General Relativity and the Standard Model.\cite{SME}
Experiments over a broad range provide striking constraints on the Lorentz-violating coefficients.\cite{datatables}
The focus of the current work is possible Penning-trap signals arising from nonminimal fermion sector 
including interactions up to mass-dimension six.

\section{Theory}
In the SME framework, 
a charged Dirac fermion $\psi$ with mass $m$ in the presence of Lorentz violation 
is described by a modified Dirac equation,\cite{nonminimal}
\begin{equation}
(iD_\mu \gamma^{\mu}-m+\hat{\mathcal{Q}})\psi=0,
\end{equation}
where $iD_\mu=i\partial_\mu-q A_\mu$, 
with $A_\mu$ being the electromagnetic four-potential.
The quantity $\hat{\mathcal{Q}}$ is a general Lorentz-violating operator 
involving covariant derivatives $iD_\mu$,
with anticommutators associated with coefficients for Lorentz violation
affecting propagation
and with commutators introducing couplings to the field
strength controlled by $F$-type coefficients for Lorentz violation. 
The study of the latter was initiated in the contexts
of noncommutative electrodynamics\cite{chklo}
and topological phases.\cite{be04} 
We note that not all the coefficients appearing in the general operator $\hat{\mathcal{Q}}$ are observables
because possible field redefinitions can be made to remove certain combinations.

For precession measurements in Penning traps,
the primary interest involves the difference between energy levels.
The trap is idealized as a uniform constant magnetic field,
and the energy shifts due to Lorentz violation are calculated using perturbation theory,
by taking the expectation value with unperturbed Landau wavefunctions 
of the perturbative hamiltonian corrected only by Lorentz violation. 
We can expect the energy shifts to contain 
terms proportional to the Lorentz-violating coefficients,
the fermion mass,
the magnetic field, 
or possible combinations. 
In a typical Penning trap,
the size of the magnetic field is of order 1-10 Tesla,
which is suppressed relative to the fermion mass by several orders of magnitude.
Therefore terms proportional to the magnetic field 
and associated with propagation effects
can be safely ignored during the analysis,
and the magnetic field plays a dominant role
only for interactions with $F$-type coefficients.

\section{Experimental signals}

There are two types of energy differences related to measurements in Penning traps,
corresponding to the cyclotron and anomaly frequencies.
Since the cyclotron motion of a fermion in a Penning trap 
is created by the presence of magnetic field in the trap,
any signals involving the cyclotron frequencies in principal 
depend on the magnetic field.
This is suppressed relative to the fermion mass,
so the cyclotron motion can be safely ignored.
Therefore in this work we focus our analysis on the experimental signals 
involving anomaly frequencies in Penning traps, 
e.g.,
studies of the $g$ factor and magnetic moment of a single fermion 
and their difference between particles and antiparticles. 

Another important feature of Lorentz violation in any local laboratory frame is the sidereal variation
due to the Earth rotation.
As a consequence the quantities measured in the laboratory oscillate in sidereal time.
We adopt the standard Sun-centered inertial frame\cite{suncenter}
to express our results.

\section{Applications and results}

Precision measurements involving particles and antiparticles in Penning traps 
can be used to set bounds 
on the Lorentz-violation coefficients.\cite{minimal,penningtrap}
Such experiments involve measurements of the ratio 
between anomaly and cyclotron frequencies.\cite{experiments}
We point out that as the transformation from the local laboratory frame to the Sun-centered frame
depends on the colatitude and the magnetic-field configuration in the trap,
so in principal each of these experiments is sensitive to different combination of coefficients.
The results are summarized in Ref.\ \refcite{penningtrap}.
They extend the range of previous work\cite{minimal} for the minimal sector 
by including the dimension-four $g$ coefficient
and also by presenting nonminimal results 
for mass dimensions five and six.

\section*{Acknowledgments}
This work was supported in part
by the Department of Energy under grant number {DE}-SC0010120
and by the Indiana University Center for Spacetime Symmetries.

\end{document}